\documentclass[ijoc,nonblindrev]{informs3} % current default for manuscript submission
\usepackage{graphicx}
\usepackage{float}
\usepackage[lofdepth,lotdepth]{subfig}
\usepackage{algorithm}
\usepackage[noend]{algpseudocode}
\usepackage{tikz}
\usetikzlibrary{shapes,arrows}
\usepackage{pifont}
\usepackage[round]{natbib}
\usepackage{amssymb}
\usepackage{lineno}
\usepackage{amsmath}
\usepackage{amsfonts}
\usepackage{booktabs}
\usepackage{textcomp}
\usepackage{hyperref}
\usepackage{multirow}
\usepackage{color}
\usepackage{natbib}
\usepackage[justification=centering]{caption}

 \bibpunct[, ]{(}{)}{,}{a}{}{,}%
 \def\bibfont{\small}%
\TheoremsNumberedThrough     % Preferred (Theorem 1, Lemma 1, Theorem 2)
\EquationsNumberedThrough    % Default: (1), (2), ...
\begin{document}
%%%%%%%%%%%%%%%%

%\setcounter{page}{1}
%\DOI{10.1287/xxxx.0000.0000}%


\RUNTITLE{Identifying intracity freight trip ends from heavy truck GPS trajectories}

% \TITLE{Bundling Information Goods of Decreasing Value}
\TITLE{Identifying intracity freight trip ends from heavy truck GPS trajectories}

\ARTICLEAUTHORS{%
\AUTHOR{Yitao Yang}
\AFF{Key Laboratory of Integrated Transport Big Data Application Technology for Transport Industry, Beijing Jiaotong University, Beijing 100044, China, \EMAIL{yitao-yang@bjtu.edu.cn}}
\AUTHOR{Bin Jia\footnote{Corresponding author}}
\AFF{Key Laboratory of Integrated Transport Big Data Application Technology for Transport Industry, Beijing Jiaotong University, Beijing 100044, China. School of Economics and Management, Xi’an Technological University, Xi’an 710021, China,  \EMAIL{bjia@bjtu.edu.cn}}
\AUTHOR{Xiao-Yong Yan\footnote{Corresponding author}}
\AFF{Key Laboratory of Integrated Transport Big Data Application Technology for Transport Industry, Beijing Jiaotong University, Beijing 100044, China, \EMAIL{yanxy@bjtu.edu.cn}}
\AUTHOR{Rui Jiang}
\AFF{Key Laboratory of Integrated Transport Big Data Application Technology for Transport Industry, Beijing Jiaotong University, Beijing 100044, China, \EMAIL{jiangrui@bjtu.edu.cn}}
\AUTHOR{Hao Ji}
\AFF{School of Economics and Management, Xi’an Technological University, Xi’an 710021, China, \EMAIL{wwwjihao\_78@126.com}}
\AUTHOR{Ziyou Gao}
\AFF{Key Laboratory of Integrated Transport Big Data Application Technology for Transport Industry, Beijing Jiaotong University, Beijing 100044, China, \EMAIL{zygao@bjtu.edu.cn}}
} % end of the block

\ABSTRACT{%
Intracity heavy truck freight trips are basic data in city freight system planning and management. In the big data era, massive heavy truck GPS trajectories can be acquired cost effectively in real-time. Identifying freight trip ends (origins and destinations) from heavy truck GPS trajectories is an outstanding problem. Although previous studies proposed a variety of trip end identification methods from different perspectives, these studies subjectively defined key threshold parameters and ignored the complex intracity heavy truck travel characteristics. Here, we propose a data-driven trip end identification method in which the speed threshold for identifying truck stops and the multilevel time thresholds for distinguishing temporary stops and freight trip ends are objectively defined. Moreover, an appropriate time threshold level is dynamically selected by considering the intracity activity patterns of heavy trucks. Furthermore, we use urban road networks and point-of-interest (POI) data to eliminate misidentified trip ends to improve method accuracy. The validation results show that the accuracy of the method we propose is 87.45\%. Our method incorporates the impact of the city freight context on truck trajectory characteristics, and its results can reflect the spatial distribution and chain patterns of intracity heavy truck freight trips, which have a wide range of practical applications.
}%


\maketitle

\section{Introduction}
The city is the center of socioeconomic development, natural resource consumption and commodity production (Acuto et al., 2018). Modern cities are supported by freight transport systems, which guarantee the supply of household goods, industrial raw materials and construction materials (Behrends, 2016). In intracity freight, heavy trucks mainly undertake mass transportation tasks between industrial enterprises, logistics warehouses and port terminals (Dernir et al., 2014). Although heavy trucks account for less than 40\% of intracity freight vehicles, they carry more than 80\% of the freight volume (Aljohani, 2016). In the future, rapid urbanization will further stimulate the growth of intracity freight demand (Balk et al., 2018). Heavy trucks will play a more important role in the intracity freight system. However, the increase in the number of heavy trucks will create serious social and environmental problems, such as traffic accidents, air pollution and nonrenewable energy consumption (Hu et al., 2019; Sakai et al., 2019; Velickovic et al., 2018). In particular, the proportion of the total mileage of heavy trucks in city road traffic is less than 6\%, but heavy trucks contribute 36\% of air pollution (Perez-Martinez et al., 2017) and 18\% of traffic death accidents (Evgenikos et al., 2016; Knight and Newton, 2008). In recent years, authorities and organizations have developed many freight-related policies, such as truck road pricing (Wang and Zhang, 2017), freight bottleneck management (Sharma et al., 2020) and freight demand management (Hassan et al., 2020), to eliminate the negative effects of heavy trucks. Intracity heavy truck freight trips are critical basic data for developing these freight policies. However, a major challenge at present is the lack of massive heavy truck freight trip data, which greatly hinders our in-depth understanding of city freight systems (Allen et al., 2012; Hadavi et al., 2019; Pluvinet et al., 2012; Zhang et al., 2019).

Traditionally, intracity freight trip data are collected through travel surveys conducted in many cities, such as London (Allen et al., 2018), Tokyo (Oka et al., 2019), Paris (Toilier et al., 2016), Toronto (McCabe et al., 2013) and Melbourne (Greaves and Figliozzi, 2008). Travel surveys provide relatively rich travel information on heavy trucks but are time consuming and costly (Allen et al., 2012; Oka et al., 2019; Pani and Sahu, 2019). Therefore, the quantity of intracity freight travel data collected through freight surveys is often limited and not sufficient for the analysis and modeling of city freight systems (Allen et al., 2014). In the era of big data, the development and application of satellite positioning technology make it possible to obtain massive heavy truck GPS trajectories through GPS devices (Deng et al., 2010; Papadopoulos et al., 2021; Pluvinet et al., 2012). It is critical to accurately extract heavy truck freight trips from GPS trajectories (Kamali et al., 2016; Zanjani et al., 2015).

To extract freight trips from GPS trajectories, most previous studies (Arentze et al., 2012; Feng et al., 2012; Gingerich et al., 2016; Huang et al., 2014; Kamali et al., 2016; Laranjeiro et al., 2019; Zanjani et al., 2015) first identified heavy truck freight trip ends (origins and destinations, OD) and then split the GPS trajectory into multiple trips according to the identified trip ends. The process of trip end identification can be divided into two steps: (1) identify truck stops from the GPS trajectory and (2) select freight trip ends from these identified truck stops. For the first step, previous studies (Comendador et al., 2012; Greaves and Figliozzi, 2008; Joubert and Axhausen, 2011; Yang et al., 2014) often used trajectory features, such as velocity, acceleration and truck direction, to infer heavy truck motion states (stationary or moving) at a given time period and then identified truck stops from GPS trajectories. These identified truck stops include not only freight trip ends but also temporary stops due to refueling, traffic congestion, etc. Thus, in the second step, freight trip ends need to be selected from all truck stops. Some researchers (Gingerich et al., 2016; Hughes et al., 2019; Zanjani et al., 2015) found that the temporary stopping time for heavy trucks is usually short (mostly a few minutes), while loading or unloading usually takes dozens of minutes or even hours. Therefore, one or more stop time thresholds can be set to distinguish freight trip ends and temporary stops.

Previous studies (Arentze et al., 2012; Feng et al., 2012; Gingerich et al., 2016; Huang et al., 2014; Laranjeiro et al., 2019) determined the time threshold for identifying freight trip ends from different perspectives. For example, some studies (Ma et al., 2011; McCormack et al., 2006; McCormack et al., 2010) took the traffic signal cycle as the time threshold to distinguish trip ends and temporary stops due to traffic control. (Kamali et al., 2016) and (Aziz et al., 2016) determined the time threshold according to city traffic conditions and the GPS data sampling rate. (Hess et al., 2015) determined the time threshold according to local freight policies. (Greaves and Figliozzi, 2008) selected the most suitable time threshold from a certain time range by a manual check. In addition, some studies (Arentze et al., 2012; Feng et al., 2012; Gingerich et al., 2016; Huang et al., 2014; Laranjeiro et al., 2019; Zanjani et al., 2015) subjectively determined different time thresholds according to the characteristics of heavy truck freight activities. Overall, the time threshold determination methods proposed in most previous studies are subjective and feasible for specific scenarios but lack universality. However, the above methods may not be suitable for intracity trip end identification since these method determined only one single time threshold. In intracity freight, heavy truck transport activities are usually organized in the form of trip chains; that is, heavy trucks start from a base (i.e., freight enterprise, logistics warehouse and factory), make multiple trips for different purposes, and finally return to this base. In a trip chain, a heavy truck dwells longer at the base and shorter at intermediate destinations. Moreover, the spatial patterns of truck trip chains are complex and diverse in different cities (Siripirote et al., 2020). Therefore, a single time threshold may not be sufficient to accurately identify heavy truck short-stay trip ends in a trip chain.

To address this issue, (Thakur et al., 2015) proposed a trip end identification method with dynamic time threshold adjustment. This method first manually determines three-time thresholds, i.e., 30 min, 15 min and 5 min, and then dynamically selects a suitable time threshold according to the circuity ratio (the ratio between the straight line distance and cumulative geodetic distance from the origin to destination) of the truck trajectory. The smaller the circuity ratio of a GPS trajectory is, the greater the likelihood that it is composed of multiple trips. This method first uses the time threshold of 30 min to identify the long-stay trip ends of heavy trucks from a GPS trajectory and then splits this trajectory into multiple segments (defined as primary subtrajectories) according to the identified trip ends. Next, this method calculates the circuity ratio of each primary subtrajectory. A subtrajectory with a circuity ratio less than a predefined threshold of 0.7 is considered to contain multiple trips due to its significantly circuitous degrees. Then, this method uses a time threshold of 15 min to identify the trip ends from each primary subtrajectory composed of multiple trips and splits these primary subtrajectories into multiple secondary subtrajectories according to the identified trip ends. Similarly, this method uses a time threshold of 5 min to identify the trip ends from each secondary subtrajectory composed of multiple trips. After this procedure, this method considers that all trip ends are identified. This time threshold dynamic adjustment method is applicable for identifying intracity heavy truck trip ends, greatly improving the identification accuracy of truck short-stay trip ends in a trip chain. However, the time thresholds (30 min, 15 min and 5 min) and circuity ratio threshold (0.7) in this method are determined subjectively. In the era of big data, massive heavy truck GPS trajectories provide the possibility to identify heavy truck trip ends objectively and accurately. However, a data-driven method of identifying intracity heavy truck trip ends is still lacking.

In this paper, we use GPS trajectories of 2.7 million heavy trucks in China as basic data and propose a data-driven method for identifying intracity heavy truck trip ends. We use a nonparametric iterative method to determine the multilevel time thresholds. In the process of trip end identification, we first select the first-level time threshold to identify trip ends from a GPS trajectory and then split this GPS trajectory into multiple subtrajectories according to the identified trip ends. Afterward, we determine whether these subtrajectories are composed of multiple trips according to the circuitous degree of intracity heavy truck trip paths. For this purpose, we use a multipath generation algorithm and a similarity analysis method to find the nth shortest path that is closest in length to the intracity heavy truck trip paths. The nth shortest path is used to measure the circuitous degree of intracity heavy truck trip paths. Similarly, we select the next level of a smaller time threshold to identify trip ends from subtrajectories composed of multiple trips. The above process is iterated until all of the short-stay heavy truck trip ends are identified. For each identified trip end, we use freight-related POIs and urban road networks to determine whether it is an actual trip end. Finally, we discuss the potential application value of extracted intracity heavy truck freight trips.

\section{Data description}
\subsection{GPS trajectory data}
Our heavy truck GPS trajectory data are from the China Road Freight Supervision and Service Platform (https://www.gghypt.net/). This platform is used to record the real-time geographic locations of all heavy trucks in China and monitor their traffic violations. We obtain the GPS trajectories of 2.7 million heavy trucks over the period from May 18, 2018 to May 24, 2018. The average sampling rate of GPS trajectory data is 30 s. The attributes of the GPS trajectory data include truck ID, timestamp, longitude, latitude, speed, and direction angle. We preprocess the raw GPS trajectory data by removing missing, duplicate and other abnormal records to improve data quality. Figure 1a-d shows the spatial distributions of preprocessed heavy truck GPS trajectories in Beijing, Chengdu, Shanghai and Suzhou in China.

\subsection{Urban road networks}
In this paper, we use urban road network as basic data to measure the circuitous degree of intracity heavy truck trip paths. We use OSMnx (Boeing, 2017), a Python package for downloading OpenStreetMap (Haklay and Weber, 2008) street network data, to obtain urban road networks. The urban road network is represented by a directed graph, in which edges refer to road segments and nodes refer to intersections. Each edge’s weight is its length. According to the freight policies of different cities (Chen et al., 2019; Wang et al., 2020a), central areas with high population density are often regulated as restricted areas for heavy trucks (see Fig. 1e-h). These restricted areas are unreachable for heavy trucks at certain time periods. Therefore, we remove the road segments within the restricted areas to construct multiple road networks accessible to heavy trucks at different time periods.

\subsection{Freight-related POI data}
In this paper, we use freight-related POI data to remove non-actual trip ends from identified trip ends to improve method accuracy. We use the application programming interface (API) of Amap (https://lbs.amap.com) to crawl freight-related POIs in each city. The attributes of each POI include name, geographic location and location type. The location types of freight-related POIs include building companies, mechanical electronics, chemical metallurgy, commercial trade, logistics warehouses, mining companies, factories, agricultural bases, industrial parks and building material markets. Figure 1i-l shows the spatial distributions of freight-related POIs in four cities. We find that a significant number of freight enterprises (the major locations where city freight traffic is generated and attracted (Gonzalez-Feliu and Sanchez-Diaz, 2019)) are located at the periphery of the city center, indicating an obvious suburbanization trend of city freight activities.

\begin{figure}
	\centering
	\includegraphics[scale=0.4]{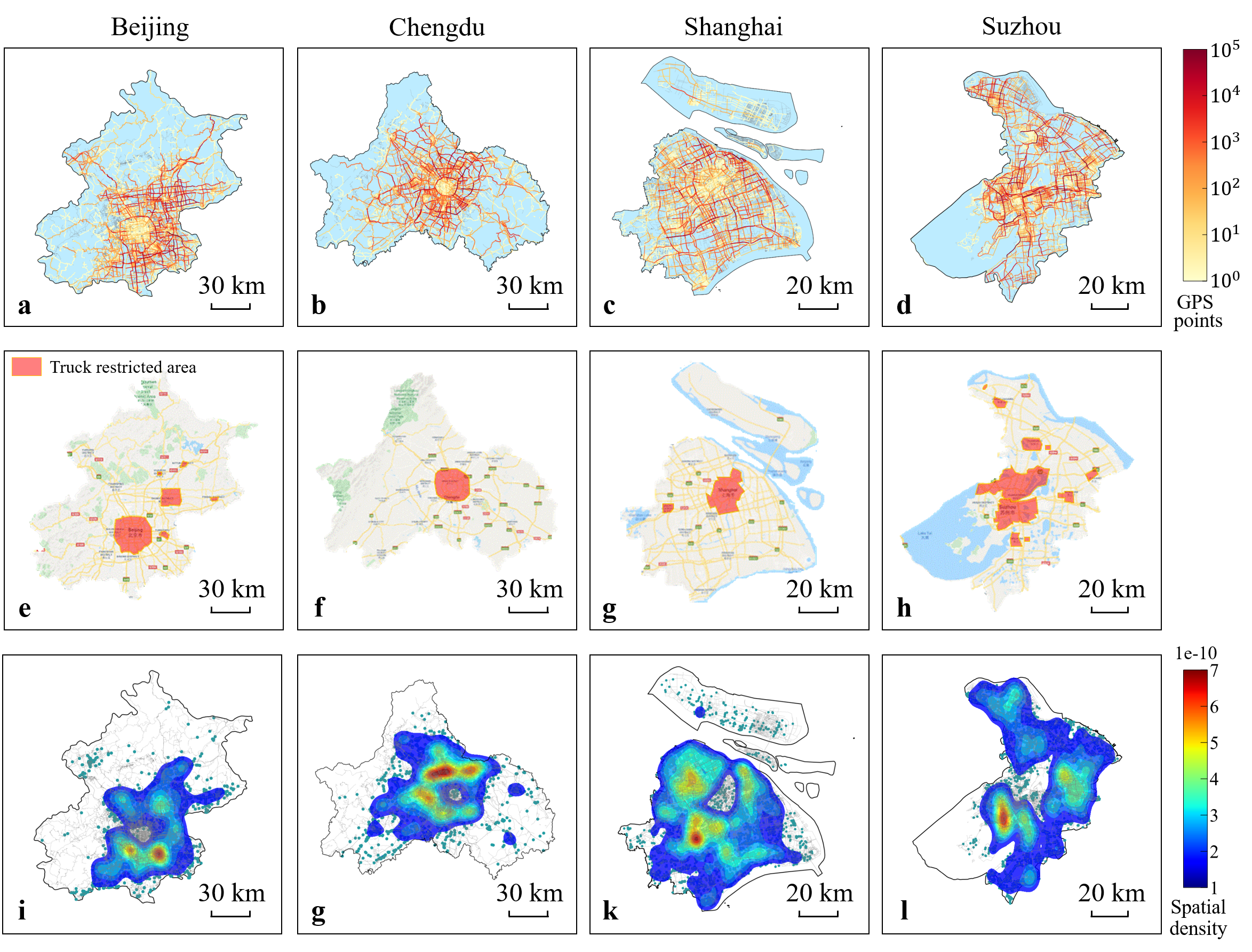}
	\captionsetup{justification=centering}
	\caption{Illustration of the main basic data. a-d Spatial distributions of heavy truck GPS trajectories in four cities: Beijing, Chengdu, Shanghai and Suzhou. The color bar represents the number of GPS trajectory points on a road segment. e-h The maximum restricted areas for heavy trucks in these four cities, whose truck restriction policies can be found at http://jtgl.beijing.gov.cn, http://cdjg.chengdu.gov.cn, https://dlysj.sh.gov.cn and http://jjzd.szgaj.cn. i-l Spatial distributions of freight-related POIs in these four cities. A green dot represents freight-related POI. The color bar represents the spatial density of freight-related POIs calculated by the kernel density estimation method (Parzen, 1962; Rosenblatt, 1956).}
\end{figure}

\section{Methodology}
We propose a data-driven method to identify the intracity trip ends of each heavy truck. Figure 2 shows the three main steps in this method. First, we identify heavy truck stops from GPS trajectories by using a predetermined speed threshold (see Fig. 2a). These identified truck stops may consist of both trip ends (loading stops, unloading stops and rest stops) and temporary stops due to refueling, traffic congestion, etc. Second, we determine multilevel time thresholds and select an appropriate time threshold level according to a truck’s maximum stopping time to identify trip ends from these truck stops. For example, the first-level (maximum) time threshold is selected if it is shorter than this truck’s maximum stopping time. Otherwise, another level of shorter time threshold is selected to ensure that trip ends are identified. Then, we split an entire GPS trajectory into multiple segments (primary subtrajectories) according to these identified trip ends (see Fig. 2b). Third, we determine whether these primary subtrajectories may be composed of multiple trips. If a primary subtrajectory is significantly circuitous, then it is likely to be composed of multiple trips, as shown in Fig. 2c. We use the next shorter time threshold level to identify potential trip ends from these significantly circuitous primary subtrajectories. If these potential trip ends are identified, one primary subtrajectory is split into multiple secondary subtrajectories, as shown in Fig. 2d. Otherwise, the next shorter time threshold level is used. The above process is iterated until there are no significantly circuitous subtrajectories or the time threshold reaches the minimum value, indicating that all trip ends in an entire trajectory are identified.

\begin{figure}
	\centering
	\includegraphics[scale=1]{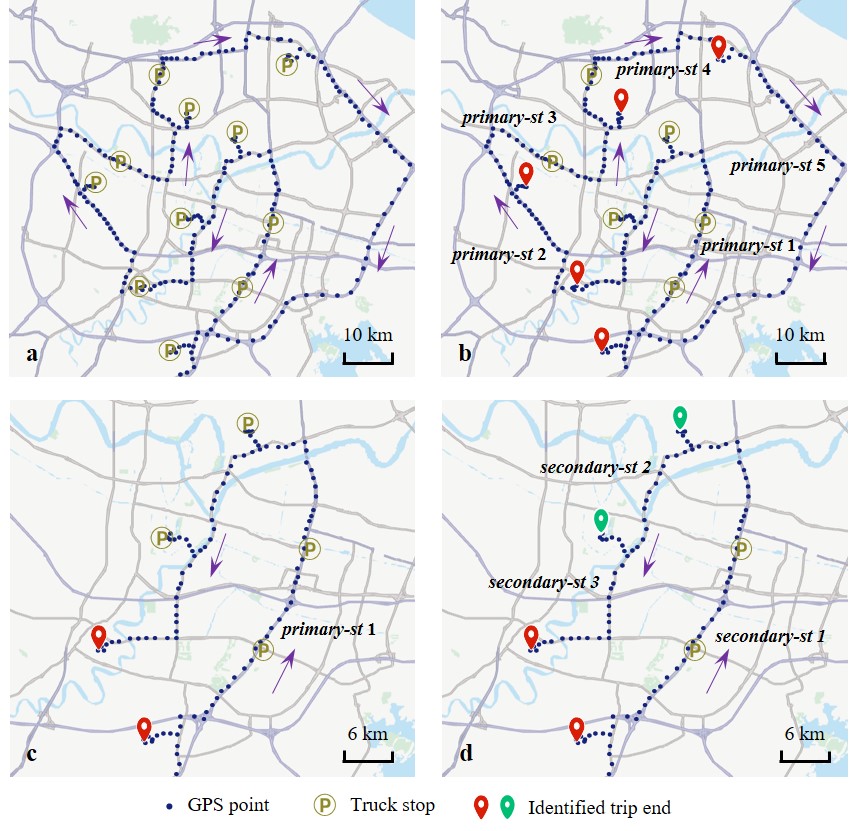}
	\captionsetup{justification=centering}
	\caption{Illustration of the proposed intracity trip end identification methodology. a Identifying truck stops from an entire GPS trajectory. The purple arrow indicates this truck's driving direction. b Selecting trip ends from truck stops by using an appropriate level of time threshold and splitting this GPS trajectory into five segments, i.e., primary subtrajectories (denoted by primary-st), according to the identified trip ends. c One primary subtrajectory may be composed of multiple trips due to its significantly circuitous degree. d Identifying potential trip ends from this primary subtrajectory by using the next one or more shorter time threshold levels. This primary subtrajectory is split into three secondary subtrajectories (denoted by secondary-st) according to the identified truck short-stay trip ends.}
\end{figure}

\subsection{Identifying truck stops from GPS trajectories}

In an entire GPS trajectory, we calculate the ratio of the geographical distance and time difference between two adjacent GPS points, i.e., truck speed in a time period, to infer the heavy truck motion state (stationary or moving) in this time period. For example, a heavy truck is considered to be stationary if its speed in a time period is zero. The location where this stationary heavy truck is located is identified as a truck stop. However, GPS data drift (Wang and Morton, 2015) may cause positioning deviation, which makes the speed value of stationary heavy trucks nonzero. Therefore, we need to determine a suitable speed threshold to identify truck stops from GPS trajectories. We use a data-driven method to define the speed threshold. In the method, we recognizes the speed characteristics induced by data drift from the distribution of heavy truck speed in each time period and select the speed value corresponding to the characteristic transition point of the speed distribution as a speed threshold.

\begin{figure}
	\centering
	\includegraphics[scale=0.28]{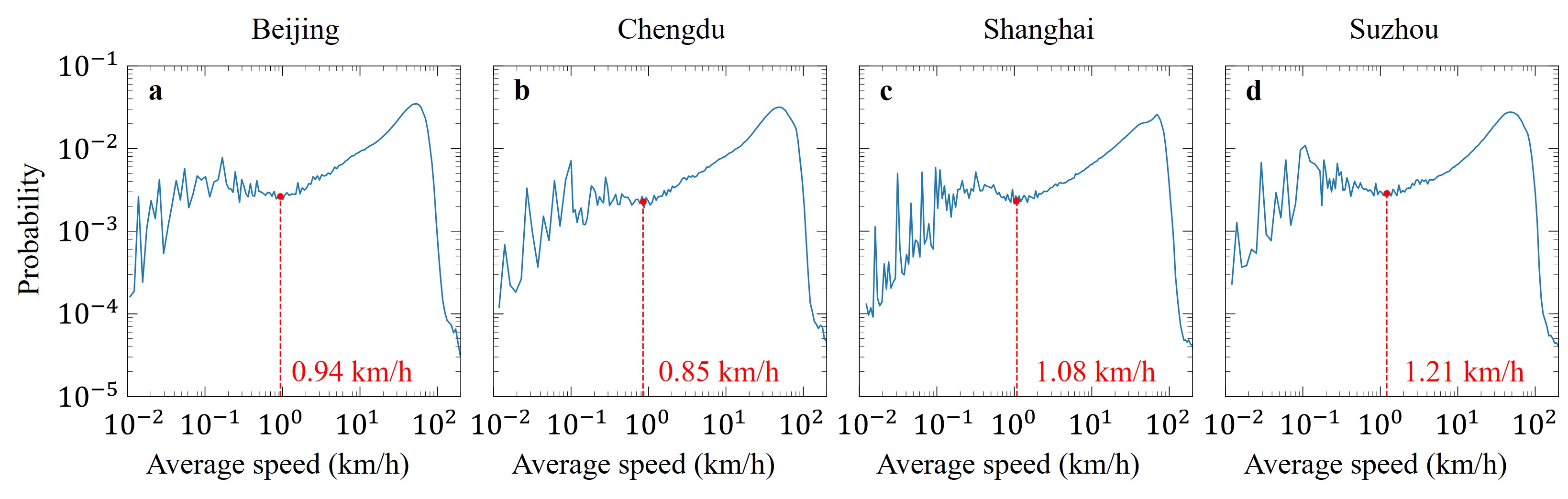}
	\captionsetup{justification=centering}
	\caption{Distributions of heavy truck speed in each time period in four cities. The red dots in panels a-d indicate the characteristic transition points of the speed distribution. The speed values corresponding to these transition points are defined as speed thresholds for these four cities.}
\end{figure}

First, we calculate the truck speed in each time period and obtain its distribution. Second, we select the speed value corresponding to the characteristic transition point of the speed distribution as a speed threshold. Figure 3a-d shows the heavy truck speed distributions in Beijing, Chengdu, Shanghai and Suzhou in China respectively. We find that the low-speed part of the distribution curve fluctuates greatly without an obvious pattern, revealing the characteristics induced by data drift. We select the characteristic transition points between low-speed part and high-speed part of the speed distribution (i.e. the local minima) as the speed threshold. In a time period, a heavy truck is considered to be stationary if its speed is less than the determined speed threshold. We take the center of the GPS points over multiple consecutive stationary time periods as an identified truck stop, as shown in Fig. 4. The geographic location of a truck stop is indicated by the average latitude and longitude of these stopped GPS points.

\begin{figure}
	\centering
	\includegraphics[scale=0.8]{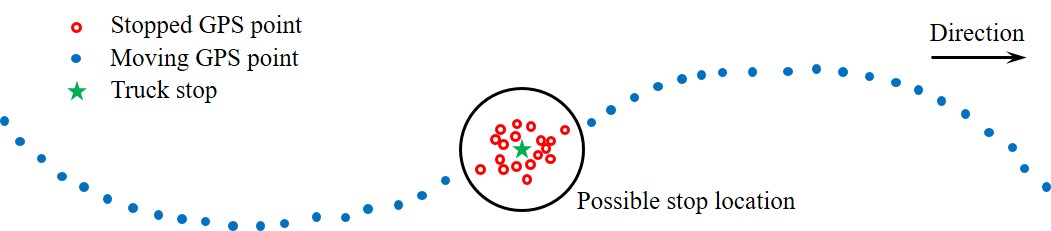}
	\captionsetup{justification=centering}
	\caption{Illustration of the geographic location of a truck stop.}
\end{figure}

\subsection{Determining multilevel time thresholds}
We use a state-of-the-art nonparametric iterative method, named the Loubar method (Bassolas et al., 2019; Louail et al., 2014), to determine multilevel time thresholds. Taking Beijing as an example, we first identify all truck stops by using the method introduced in Section 3.1. Second, we sort the truck stops in descending order of stopping time and calculate the cumulative stopping time of heavy trucks at the sorted stops. Finally, we normalize the ordinal number of sorted truck stops and the corresponding cumulative stopping time to plot the Lorenz curve (Lorenz, 1905), as shown in Fig. 5a. The curvature value of the Lorenz curve indicates the equilibrium degree of the data distribution. A larger curvature shows a greater number of truck long-stay stops. According to the Loubar method, we calculate the intersection point F* of the tangent line of the Lorenz curve at (1, 1) with the horizontal axis. The truck long-stay stops with normalized ordinal numbers greater than F* are divided into the first class. The stopping time of a heavy truck at a stop corresponding to F* is determined as the first-level time threshold, as shown in Fig. 5a. Next, we remove these truck long-stay stops, then redraw the Lorenz curve for the remaining truck stops to calculate the corresponding F* and determine the next time threshold level, as shown in Fig. 5b. The above process is repeated (see Fig. 5c-g) until a balanced data distribution is reached, i.e., the Lorenz curve is a straight line, as shown in Fig. 5h. Finally, we determine 7 time threshold levels, i.e., 1,295 min, 326 min, 72 min, 26 min, 12 min, 2 min and 1 min. Statistical analysis shows that the maximum stopping time of each heavy truck is greater than the seventh-level time threshold of 1 min and that of 80\% of the heavy trucks is less than the first-level time threshold of 1,295 min, which indicates that all heavy truck trip ends can be identified by selecting appropriate time threshold levels according to the maximum stopping time of this heavy truck. For different cities, respective multilevel time thresholds are determined by using the above method.

\begin{figure}
	\centering
	\includegraphics[scale=0.35]{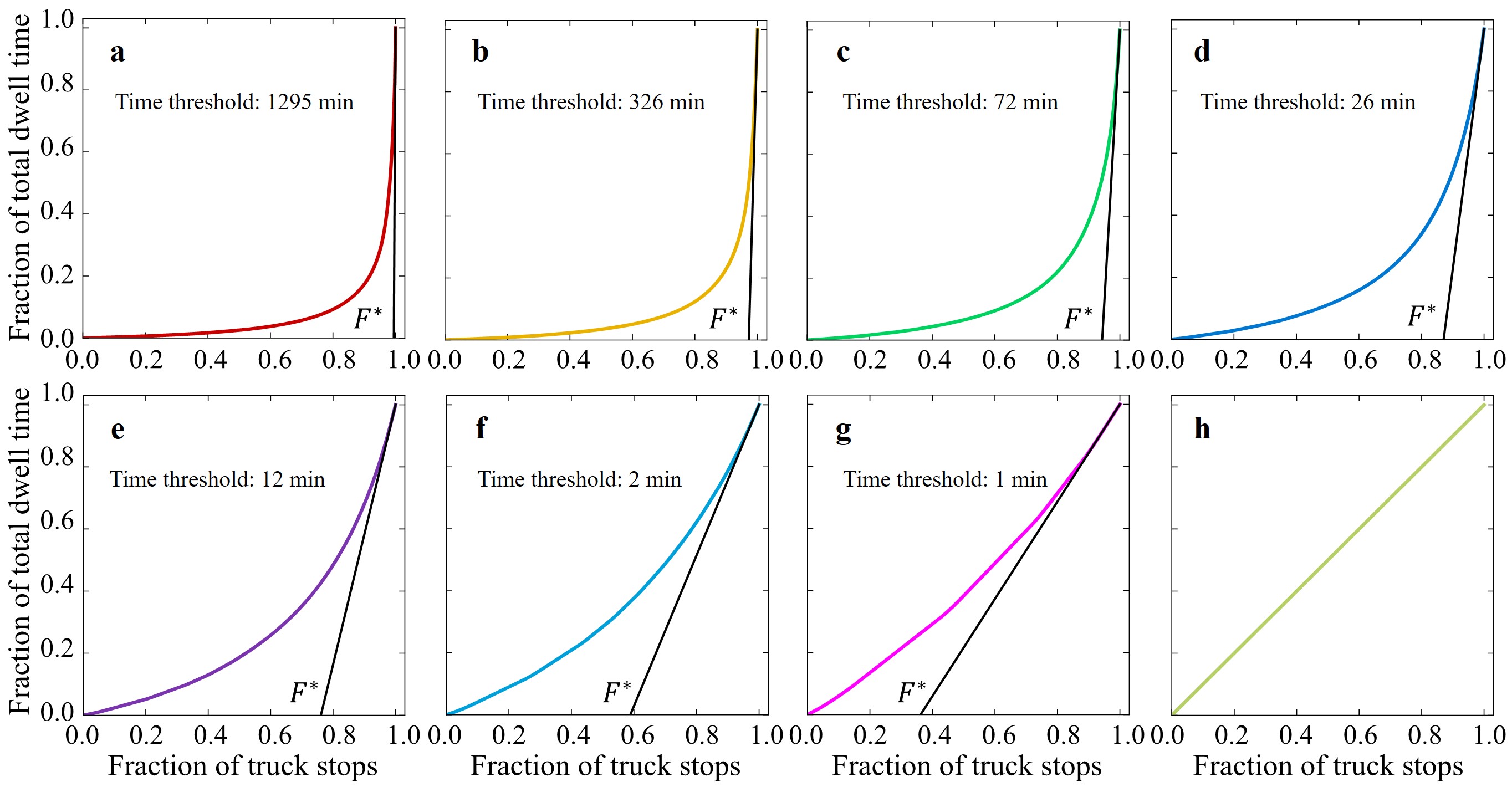}
	\captionsetup{justification=centering}
	\caption{Multilevel time thresholds were determined by using the Loubar method in Beijing. a Lorenz curve of heavy truck stopping time at all stops. The red curve represents the Lorenz curve. The black line represents the tangent Lorenz curve at (1, 1), and F{$^*$} is the intersection point between this tangent and the horizontal axis. The heavy truck stopping time at the stop corresponding to F* is determined as a time threshold. b The Lorenz curve of heavy truck stopping time shorter than the time threshold determined in the previous panel, i.e., panel a. and so on for panels c-h.}
\end{figure}

\subsection{Identifying potential trip ends from circuitous subtrajectories}
In the process of intracity trip end identification, we first select an appropriate time threshold level according to the maximum stopping time of a heavy truck to identify trip ends from an entire GPS trajectory and then split this trajectory into multiple subtrajectories. Each subtrajectory may be composed of one trip or multiple trips. Next, we need to identify the subtrajectory that may be composed of multiple trips and then use the next level of the shorter time threshold to identify potential short-stay trip ends.

Here, we determine whether a subtrajectory may be composed of multiple trips according to its circuitous degree. The city freight context, as indicated by traffic conditions, truck restriction policies, etc., affects the circuitous degree of heavy truck trajectories. In a single trip, a heavy truck driver may choose a longer path than the shortest path to bypass truck restricted areas (Kamali et al., 2016; Luong et al., 2018; Oka et al., 2019), so a truck single trip path is inevitably circuitous. In a subtrajectory composed of multiple trips, a heavy truck needs to serve several clients at different locations, so the circuitous degree of its path tends to be greater than that of a single trip path (Duan et al., 2020a; Moshref-Javadi et al., 2020; Sakai et al., 2017; Zhen et al., 2020). Therefore, we need to measure the circuitous degree of the truck single trip path and use this to determine whether a subtrajectory may be composed of multiple trips. First, we use satellite images to conduct geospatial analysis (Duan et al., 2020b; Gingerich et al., 2016) on truck GPS trajectories to manually extract some heavy truck single trips. Second, we use a multipath generation algorithm (Leblanc et al., 1975) to generate the K shortest paths from the origin to the destination of each single trip. Figure 6 shows the path of a single trip (denoted by the yellow line) and the corresponding eight shortest paths. Third, we compare the length of each single trip path with that of the corresponding K shortest paths and then use proximity analysis (Sorensen, 1948) to identify the nth (n{$<$}=K) shortest paths that are closest in length to these single trip paths overall. The nth shortest path corresponding to a single trip is used to measure the circuitous degree of the path of this single trip. For a split subtrajectory, we first generate the nth shortest path from its origin to destination and then compare the length of the nth shortest path with that of the actual path. This subtrajectory may be composed of multiple trips if its length is longer than the corresponding nth shortest path.

\begin{figure}
	\centering
	\includegraphics[scale=0.6]{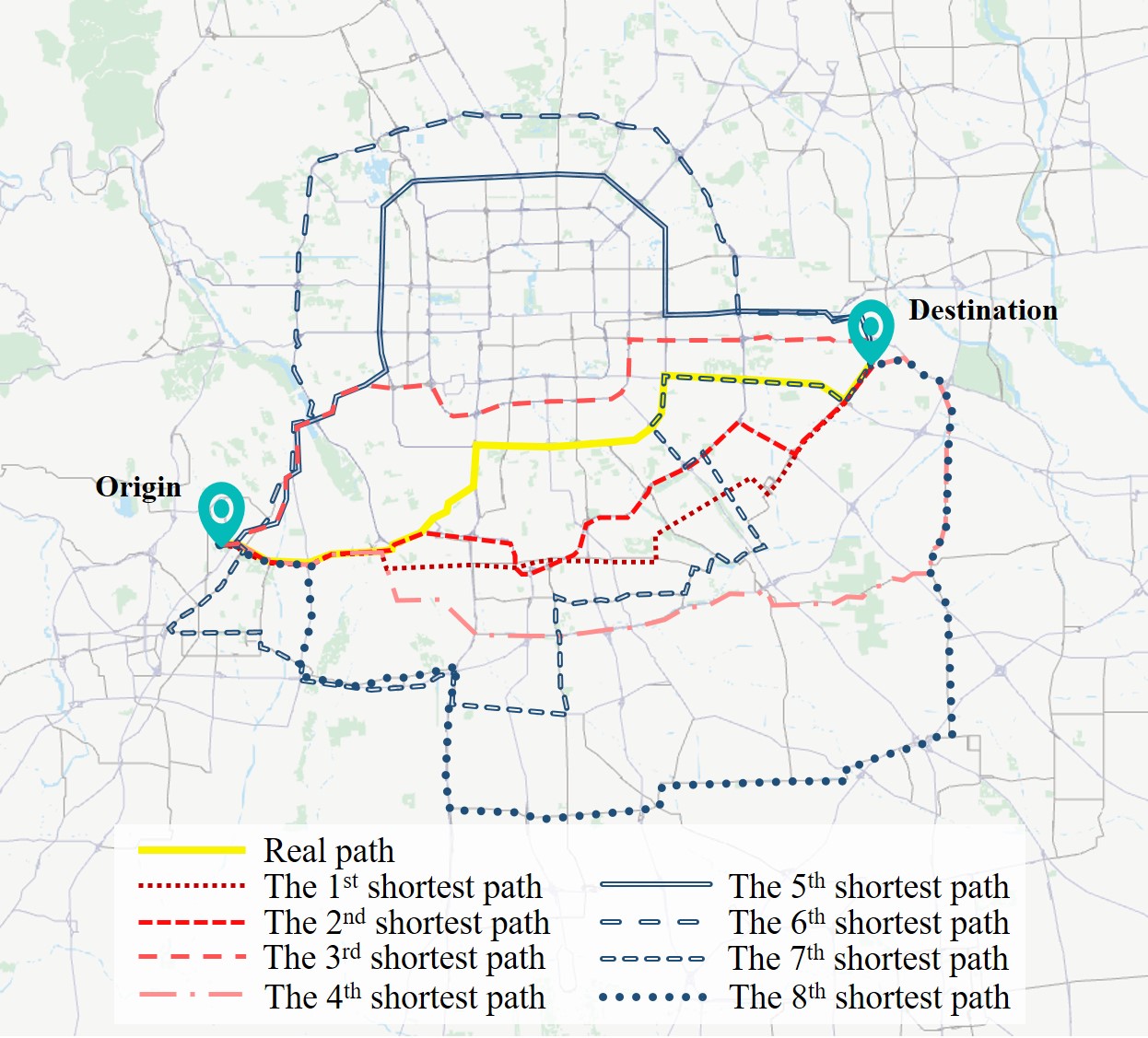}
	\captionsetup{justification=centering}
	\caption{Multilevel time thresholds were determined by using the Loubar method in Beijing. a Lorenz curve of heavy truck stopping time at all stops. The red curve represents the Lorenz curve. The black line represents the tangent Lorenz curve at (1, 1), and F{$^*$} is the intersection point between this tangent and the horizontal axis. The heavy truck stopping time at the stop corresponding to F* is determined as a time threshold. b The Lorenz curve of heavy truck stopping time shorter than the time threshold determined in the previous panel, i.e., panel a. and so on for panels c-h.}
\end{figure}

Here we present a case study for Beijing. First, we manually extract nearly 2,000 truck single trips from GPS trajectories by using geospatial analysis. Since some of the loading or unloading locations have distinct architectural features that can be recognized from satellite maps, we can conduct geospatial analysis to extract some single trips with these locations as origins and destinations. Second, we use the link elimination method (Leblanc et al., 1975) to generate the K shortest paths from the origin to the destination of each single trip. In this method, we use the Dijkstra algorithm (Dijkstra, 1959) to obtain the shortest path for a given pair of origin and destination, and then remove the road segment located in the middle of that path before searching for the next shortest path. Finally, we use the Sørensen similarity index (SSI) (Sorensen, 1948) to find the nth shortest paths that are closest in length to all single trip paths overall. Figure 7 shows that the 3rd shortest path (with the maximum SSI value and R\^2) is applicable to measure the circuitous degree of a heavy truck single trip path in Beijing. For a split subtrajectory, we generate the 3rd shortest path from its origin to destination in the urban road network. This subtrajectory may be composed of multiple trips if its length is longer than the corresponding 3rd shortest path. Then, the next shorter time threshold level is used to identify potential trip ends from this subtrajectory.

\begin{figure}
	\centering
	\includegraphics[scale=0.35]{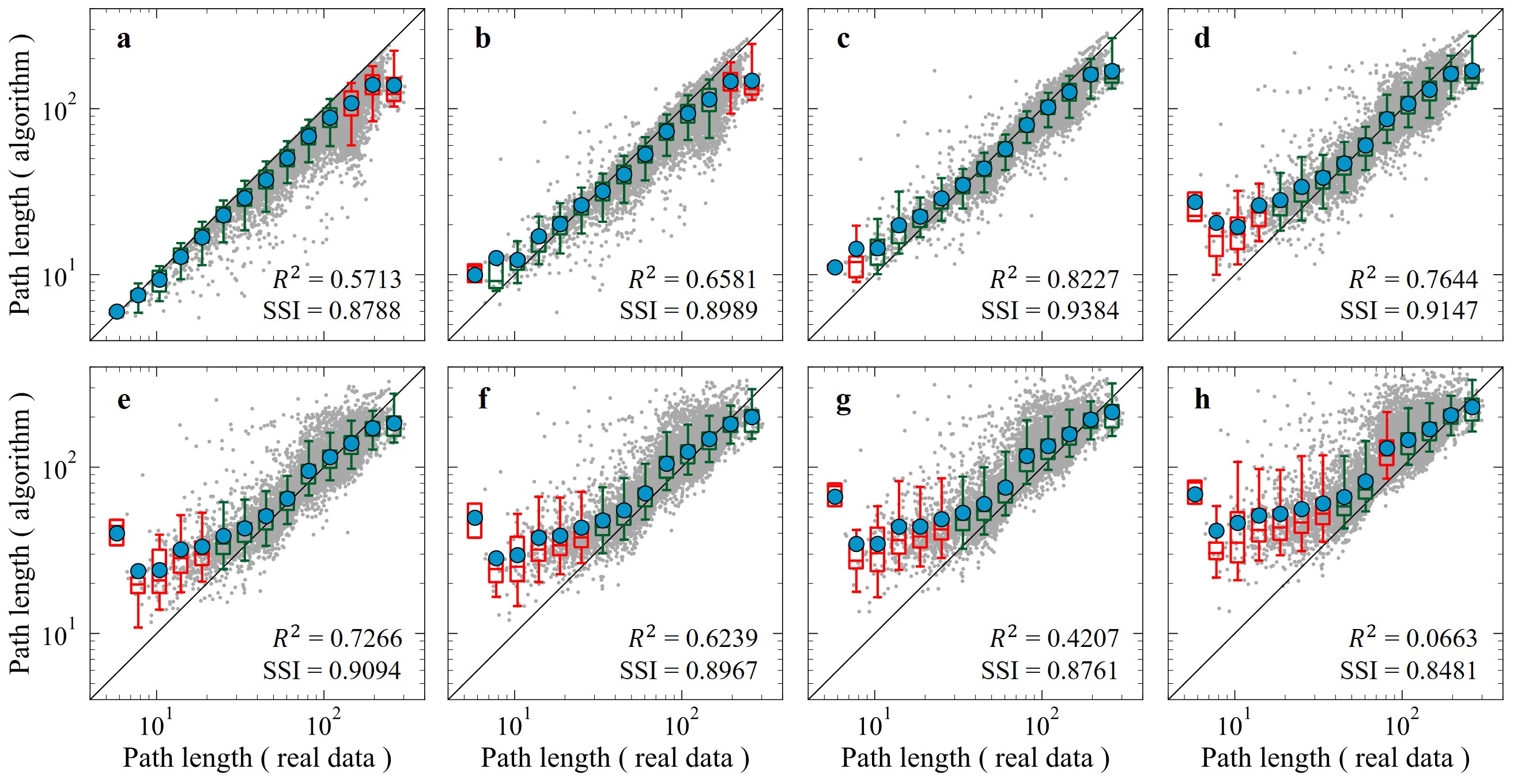}
	\captionsetup{justification=centering}
	\caption{Measuring length proximity between actual paths and the corresponding eight shortest paths in Beijing. a-h display of the proximity measure results between actual paths and the first to eighth shortest paths. The gray points are scatter plots for each trip. The standard boxplot represents the distribution of the length of the K shortest paths in different bins of the length of actual paths. The blue points represent the average length of the K shortest paths in different bins. A box is marked in green if the line y = x lies between 10\% and 91\% in that bin and is marked in red otherwise.}
\end{figure}

\subsection{Eliminating misidentified trip ends}
In this paper, we identify heavy truck trip ends from GPS trajectories by using multilevel time thresholds. In practice, it is notable that the duration of heavy truck temporary stops may be similar to the time required for loading or unloading. In this case, some temporary stops may be misidentified as trip ends. Hence, we need to eliminate misidentified trip ends by using urban road networks and freight-related POI data to improve method accuracy.

First, we use urban road networks to eliminate misidentified trip ends located on the road. Urban road networks consist of four road classes: motorways, primary roads, secondary roads and tertiary roads. We obtain the average width of each of the four road classes according to Chinese road construction standards and set it as the buffer distance. If the minimum distance between an identified trip end and the centerline of the nearest road is less than half of the buffer distance, this trip end is misidentified and needs to be eliminated, as shown in Fig. 8a. Otherwise, we use freight-related POI data to determine whether this trip end is accurately identified. If there is a freight-related POI near the trip end, this trip end is a real end, as shown in Fig. 8b, and vice versa, as shown in Fig. 8b.

\begin{figure}
	\centering
	\includegraphics[scale=0.5]{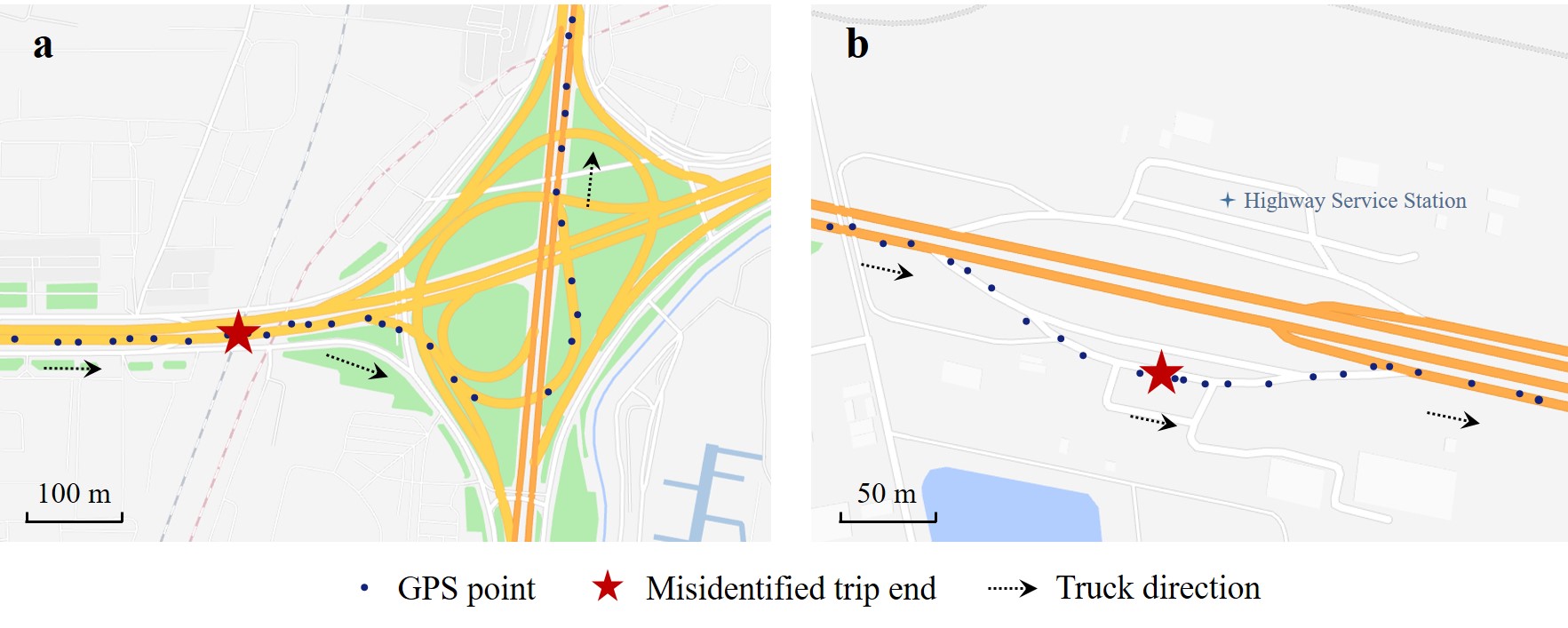}
	\captionsetup{justification=centering}
	\caption{Illustration of misidentified trip ends. a A misidentified trip end is located on the road. This heavy truck may be stuck on the road for a long time due to traffic congestion, traffic control, etc. b A misidentified trip end not located on the road has no freight-related POI. This heavy truck may stop at a highway service station because the driver is on a temporary break.}
\end{figure}

\subsection{Method validation}

In this paper, we use satellite images and freight-related POI data to conduct geographic analysis (Gingerich et al., 2016) to validate our proposed method. We randomly extract 1,000 heavy trucks in Beijing as a sample for method validation. Figure 9a shows the validation results of trip end identification for a typical heavy truck. The four-sided shape represents the accurately identified trip end. The three-sided shape represents the misidentified trip end. The five-sided shape represents the real trip end that is incorrectly eliminated because there is no freight-related POI near it. Figure 9b shows an accurately identified trip end that is located at a freight enterprise (not located on the road) with a freight-related POI. Figure 9c shows a case of a trip end being misidentified. In this case, the identified trip end is not located on the road, and there is a freight-related POI near it. We can find that this heavy truck stopped at a gas station from the satellite map. Hence, we assume this trip end is misidentified. Figure 9d shows that a real trip end is wrongly eliminated. This real trip end is located at a factory from the satellite image, indicating that the heavy truck is conducting freight activities, such as loading or unloading. However, our method incorrectly eliminated this real trip end because this factory has no freight-related POI, as detailed in Section 3.4. According to the above analysis, the method accuracy is calculated as:

\begin{flushright}
	$M_{acc}$=NA⁄((NA+NM+NE)),\qquad\qquad\qquad\qquad\qquad\quad(1)
\end{flushright} 

where $M_{acc}$ denotes method accuracy. NA, NM and NE denote the number of accurately identified trip ends, misidentified trip ends and incorrectly eliminated real trip ends, respectively. The validation results indicate that the method accuracy is 87.45\% and that the percentages of NM and NE are 10.67\% and 1.88\%, respectively.

\begin{figure}
	\centering
	\includegraphics[scale=0.7]{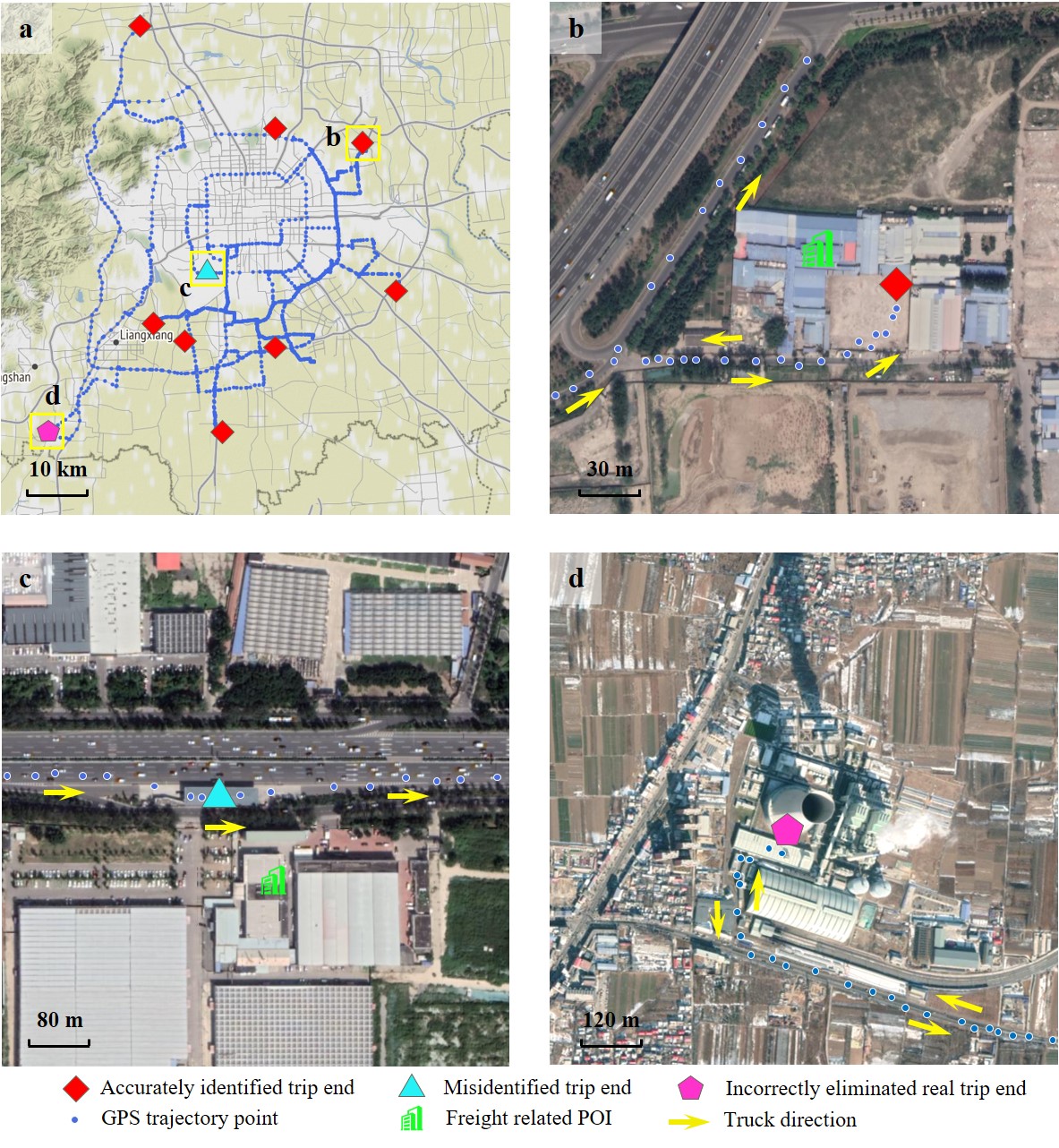}
	\captionsetup{justification=centering}
	\caption{Method validation for one typical heavy truck in Beijing. A The results of trip end identification and validation. A dot denotes a GPS point of this heavy truck in one week. The three locations marked by rectangular boxes correspond to locations in panels b-d. b A case in which a trip end is accurately identified. This trip end is located at a freight enterprise with freight-related POIs. c A case of a trip end being misidentified. This trip end is located at a gas station, not at the nearby freight enterprise. d A case in which a real trip end is incorrectly eliminated. This trip end is located at a factory with no freight-related POI.}
\end{figure}

\section{Results and analysis}
\subsection{Results of intracity freight trip end identification}
We select Beijing, Chengdu, Shanghai and Suzhou as case cities and use the above method to identify intracity heavy truck trip ends from GPS trajectories. The datasets of these four cities contain 64,000, 62,000, 94,000 and 92,000 heavy trucks, and the recording duration of the GPS trajectory is one week. First, we extract intracity heavy truck GPS trajectories by using city administrative division data, which are downloaded from OpenStreetMap. Second, we determine a speed threshold according to the truck speed distribution (see Section 3.1) and then identify truck stops from GPS trajectories by using this speed threshold. Third, we use the Loubar method to determine multilevel time thresholds. Fourth, we dynamically select appropriate time threshold levels according to the circuitous degree of a heavy truck’s single trip path to identify trip ends from truck stops. Fifth, we use freight-related POIs and urban road networks to determine whether each identified trip end is a real end to improve method accuracy. Finally, we identify 611,000, 659,000, 1,082,000 and 885,000 heavy truck trip ends in the four cities.

\subsection{Results of intracity freight trip end identification}
Spatial distributions of heavy truck trip ends can be used to recognize city freight hotspots. Figure 10 shows the city freight hotspots of four case cities, from which we can see that most freight hotspots are concentrated in suburban areas, reflecting the suburbanization trend of heavy truck freight activities (Cidell, 2010). The accurate identification of city freight hotspots can provide supporting information for authorities to effectively manage freight enterprises (Bao et al., 2019), monitor environmental pollution (Kovac et al., 2020) and plan land use (Stevens et al., 2020).

\begin{figure}
	\centering
	\includegraphics[scale=0.7]{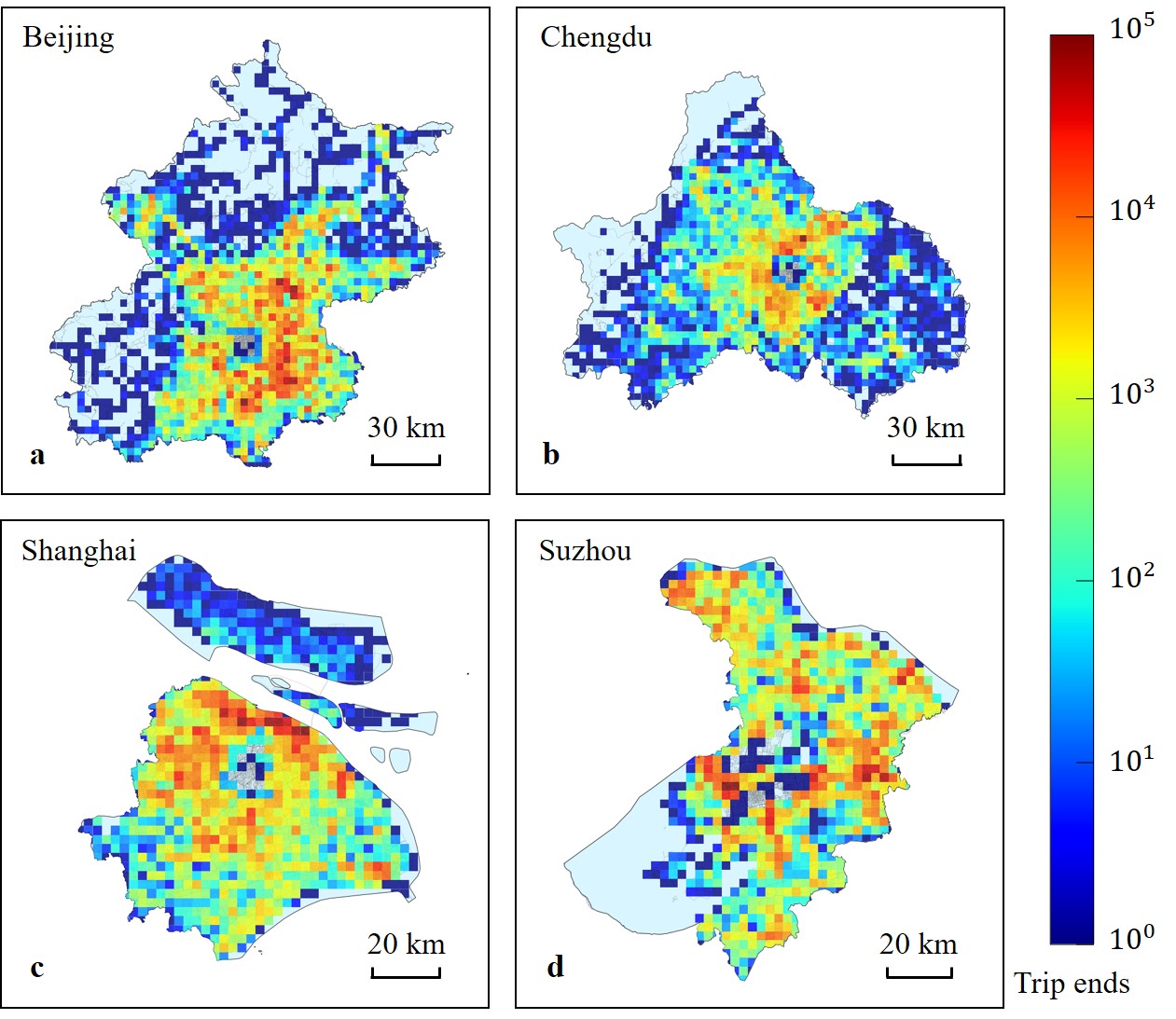}
	\captionsetup{justification=centering}
	\caption{Method validation for one typical heavy truck in Beijing. A The results of trip end identification and validation. A dot denotes a GPS point of this heavy truck in one week. The three locations marked by rectangular boxes correspond to locations in panels b-d. b A case in which a trip end is accurately identified. This trip end is located at a freight enterprise with freight-related POIs. c A case of a trip end being misidentified. This trip end is located at a gas station, not at the nearby freight enterprise. d A case in which a real trip end is incorrectly eliminated. This trip end is located at a factory with no freight-related POI.}
\end{figure}

\subsection{Spatial distribution of intracity freight trips}
We further extract the freight trip sequence of each heavy truck from its identified trip ends in a week. One freight trip can be extracted according to two consecutive trip ends, i.e., the former trip end is the origin of this trip and the latter one is the destination. In total, we extract 547,000, 597,000, 988,000 and 793,000 heavy truck freight trips in Beijing, Chengdu, Shanghai and Suzhou, respectively. Figure 11 shows the spatial distributions of freight trips in the four cities. We find that freight interactions between freight hotspots are significantly greater than those between other city areas. From the perspective of city planning and management, these extracted freight trips can be used to analyze the economic linkages and freight interaction intensity between city areas, thus providing a reference for freight policy making (Le Pira et al., 2017; Wang and Yang, 2018). In addition, heavy truck trip data can be used to calculate freight traffic generation and attraction, which are the basic data in city freight transportation planning (Jiang et al., 2020; Wang et al., 2019; Zhao et al., 2020). From the perspective of corporate operations management, heavy truck trip data can be used to estimate freight costs (Perez-Martinez and Vassallo-Magro, 2013) and freight demand (Hassan et al., 2020), which are crucial for transportation cost control and truck scheduling. From the perspective of freight network optimization, heavy truck trip data can be used to identify freight traffic bottlenecks (Kocatepe et al., 2020; McCormack et al., 2012), pointing the way to improve freight network performance.

\begin{figure}
	\centering
	\includegraphics[scale=0.7]{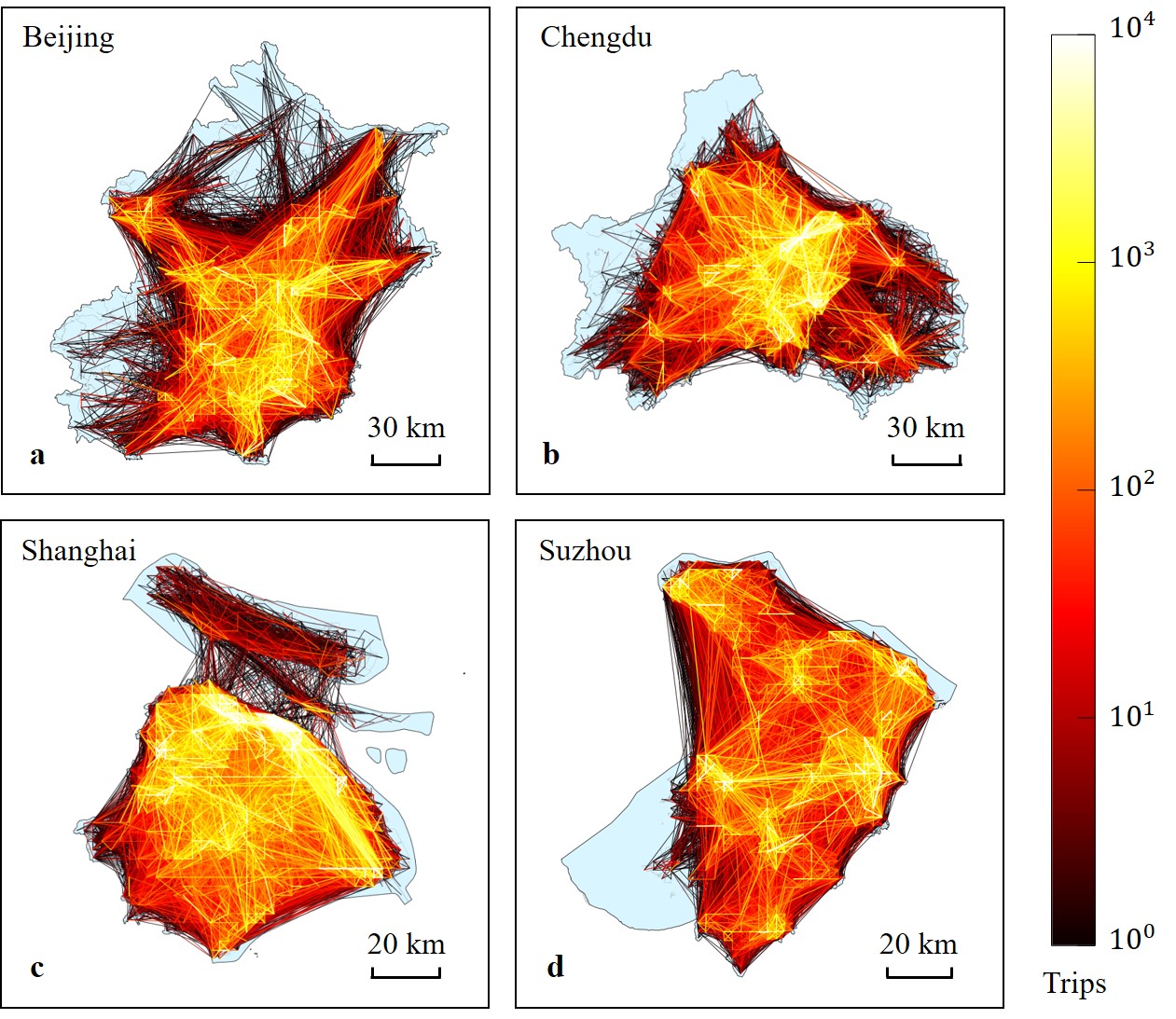}
	\captionsetup{justification=centering}
	\caption{ Spatial distributions of intracity freight trips in four cities. We partition all cities into equal-area square zones, each of which is of dimension 3 × 3 km. The color bar represents the number of freight trips between two zones.}
\end{figure}

\subsection{Analysis of intracity heavy truck trip chain patterns}
Here, we extract intracity trip chains from the trip sequence of each heavy truck. Intracity heavy truck freight activities are normally conducted in the form of trip chains (liu et al., 2021; Siripirote et al., 2020), the patterns of which are critical for understanding heavy truck freight activities. Here, we use the travel network analysis method (Yan et al., 2017) to identify typical freight trip chains from the trip sequence of each heavy truck. First, we use the density-based spatial clustering of applications with noise (DBSCAN) algorithm (Ester et al., 1996) to cluster spatially adjacent trip ends of a heavy truck into one single point, which is one node of the travel network of this heavy truck. Second, we take the trips between two nodes as directed edges of this heavy truck travel network. A heavy truck travels on this network according to its trip sequence, and the node most visited by this heavy truck is considered as the base point. We split a trip sequence into multiple trip chains according to the base point, which is the endpoint of each trip chain. In total, we identify 130 heavy truck trip chain patterns in Beijing, Chengdu, Shanghai and Suzhou. Figure 12 shows the top 10 typical trip chain patterns with the highest proportion in these four cities. We find that the proportions of various typical trip chain patterns in these four cities are similar. In Fig. 12, we can see that the trip chain pattern containing one single intermediate destination (not a base point) has the highest proportion (greater than 60\%), while the trip chain patterns containing multiple intermediate destinations only account for 40\%. Commonly, the transportation efficiency of heavy trucks in a trip chain containing multiple intermediate destinations is higher than that containing one single intermediate destination (Gonzalez-Calderon et al., 2021). Hence, authorities need to focus on truck scheduling optimization (Phan and Kim, 2016) to increase the proportion of trip chains, including multiple intermediate destinations, to improve heavy truck transportation efficiency.

\begin{figure}
	\centering
	\includegraphics[scale=0.85]{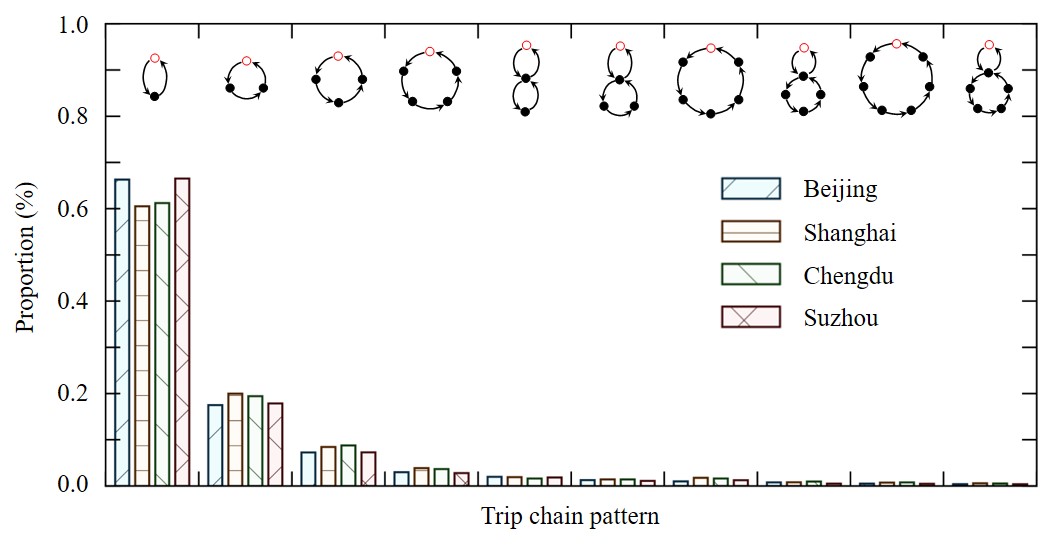}
	\captionsetup{justification=centering}
	\caption{Proportion of heavy truck trip chain patterns in four cities. The top 10 graphs show the typical trip chain patterns. The hollow circle represents the truck base point, and the solid circle represents one intermediate destination in a trip chain.}
\end{figure}

\section{Conclusion and discussion}
Extracting freight trips from massive GPS trajectories is of vital importance for city freight system planning, freight traffic management and corporate operation. Here, we propose a data-driven method to identify intracity heavy truck trip ends. In the method, the key threshold parameters are defined objectively, and the intracity heavy truck travel characteristics are considered in the process of trip end identification. First, we determine the speed threshold by analyzing the speed distribution of heavy trucks in a city to identify truck stops from GPS trajectories. Second, we determine multilevel time thresholds by using a nonparametric iterative method and dynamically select appropriate levels of time thresholds to identify trip ends considering the circuitous characteristics of intracity heavy truck trajectories. Furthermore, we use urban road networks and freight-related POI data to determine whether each identified trip end is a real trip and eliminate misidentified trips. Finally, we use satellite images and freight-related POI data to conduct method validation, the results of which suggest that intracity heavy truck freight trip ends can be accurately identified from GPS trajectories by using the method we proposed.

We apply the above method to identify heavy truck trip ends in four Chinese cities, i.e., Beijing, Chengdu, Shanghai and Suzhou. First, we recognize freight hotspots in these four cities by using the spatial distributions of heavy truck trip ends and analyze the suburbanization trend of freight activities. Second, we extract intracity heavy truck freight trips according to the identified trip ends and analyze the freight interactions between different urban areas. Finally, we use the travel network analysis method to identify typical trip chain patterns of heavy trucks in the above four Chinese cities. These results can provide supporting information for freight demand management, land use planning, freight transportation network optimization and efficiency improvement.

In addition to the potential fields of freight trip data discussed above, heavy truck freight trips can also be used to study the hierarchical organization and sustainable sprawl of cities. In recent years, rapid urbanization has raised concerns among authorities and scholars about city livability and sustainability (Sun et al., 2020). City sustainable development is facing an imbalance in transportation supply and demand at present, which weakens the livability and economic dynamics in cities (Wang et al., 2020b). Massive heavy truck freight trips can be used to understand the spatial distribution of freight demand and to analyze the constraints of freight supply. The results can be used to improve city livability and freight sustainability.

Although the accuracy of our proposed method is remarkable, there is still potential for further improvement. For example, if we can obtain heavy truck status information (e.g., engine start/shutdown) from onboard units (OBUs), we can use it to identify truck stops more accurately. In addition, some of the heavy truck loading and unloading locations have no freight-related POIs, which may lead to the real trip ends of heavy trucks being incorrectly eliminated. Therefore, richer freight-related POI data obtained by remote sensing (Shi et al., 2020; Yang et al., 2019) provide the possibility to further improve method accuracy.

\section{Acknowledgements}
This work was supported by the National Key R\&D Program of China (No. 2018YFB1600900) and the National Natural Science Foundation of China (Nos. 71621001, 71822102, 71971015).

\section{Reference}

Acuto, M., Parnell, S., Seto, K.C., 2018. Building a global urban science. Nature Sustainability 1(1), 2-4. doi:10.1038/s41893-017-0013-9.

Aljohani, K., 2016. Integrating logistics facilities in Inner Melbourne to alleviate impacts of urban freight transport. 15p. 

Allen, J., Ambrosini, C., Browne, M., Patier, D., Routhier, J.-L., Woodburn, A., 2014. Data Collection for Understanding Urban Goods Movement, In: Gonzalez-Feliu, J., Semet, F., Routhier, J.-L. (Eds.), Sustainable Urban Logistics: Concepts, Methods and Information Systems. Springer Berlin Heidelberg, Berlin, Heidelberg, pp. 71-89. doi:10.1007/978-3-642-31788-0\_5.

Allen, J., Browne, M., Cherrett, T., 2012. Survey Techniques in Urban Freight Transport Studies. Transport Reviews 32(3), 287-311. doi:10.1080/01441647.2012.665949.

Allen, J., Piecyk, M., Piotrowska, M., McLeod, F., Cherrett, T., Ghali, K., Nguyen, T., Bektas, T., Bates, O., Friday, A., Wise, S., Austwick, M., 2018. Understanding the impact of e-commerce on last-mile light goods vehicle activity in urban areas: The case of London. Transportation Research Part D-Transport and Environment 61, 325-338. doi:10.1016/j.trd.2017.07.020.

Arentze, T., Feng, T., Robroeks, J., van Brakel, M., Huibers, R., 2012. Compliance with and influence of a new in-car navigation system for trucks: Results of a field test. Transport Policy 23, 42-49. doi:10.1016/j.tranpol.2012.06.011.

Aziz, R., Kedia, M., Dan, S., Basu, S., Sarkar, S., Mitra, S., Mitra, P., 2016. Identifying and Characterizing Truck Stops from GPS Data, In: Perner, P. (Ed.), Advances in Data Mining: Applications and Theoretical Aspects, pp. 168-182. doi:10.1007/978-3-319-41561-1\_13.

Balk, D., Leyk, S., Jones, B., Montgomery, M.R., Clark, A., 2018. Understanding urbanization: A study of census and satellite-derived urban classes in the United States, 1990-2010. Plos One 13(12). doi:10.1371/journal.pone.0208487.

Bao, X., Xing, X., Zhang, D., 2019. Research on Freight Pricing Mechanism of Shipping Companies Considering Supply Chain Management. Journal of Coastal Research, 568-572. doi:10.2112/si94-112.1.

Bassolas, A., Barbosa-Filho, H., Dickinson, B., Dotiwalla, X., Eastham, P., Gallotti, R., Ghoshal, G., Gipson, B., Hazarie, S.A., Kautz, H., Kucuktunc, O., Lieber, A., Sadilek, A., Ramasco, J.J., 2019. Hierarchical organization of urban mobility and its connection with city livability. Nature Communications 10. doi:10.1038/s41467-019-12809-y.

Behrends, S., 2016. Recent developments in urban logistics research - a review of the proceedings of the International Conference on City Logistics 2009-2013, In: Taniguchi, E., Thompson, R.G. (Eds.), Ninth International Conference on City Logistics, pp. 278-287. doi:10.1016/j.trpro.2016.02.065.

Boeing, G., 2017. OSMnx: New methods for acquiring, constructing, analyzing, and visualizing complex street networks. Computers Environment and Urban Systems 65, 126-139. doi:10.1016/j.compenvurbsys.2017.05.004.

Chen, X., Wu, G., Li, D., 2019. Efficiency measure on the truck restriction policy in China: A non-radial data envelopment model. Transportation Research Part A-Policy and Practice 129, 140-154. doi:10.1016/j.tra.2019.08.010.

Cidell, J., 2010. Concentration and decentralization: the new geography of freight distribution in US metropolitan areas. Journal of Transport Geography 18(3), 363-371. doi:10.1016/j.jtrangeo.2009.06.017.

Comendador, J., López-Lambas, M.E., Monzón, A., 2012. A GPS Analysis for Urban Freight Distribution. Procedia \- Social and Behavioral Sciences 39, 521-533. doi:10.1016/j.sbspro.2012.03.127.

Deng, Z., Ji, M., American Society of Civil, E., 2010. Deriving Rules for Trip Purpose Identification from GPS Travel Survey Data and Land Use Data: A Machine Learning Approach. pp 768-777. 

Dernir, E., Bektas, T., Laporte, G., 2014. A review of recent research on green road freight transportation. European Journal of Operational Research 237(3), 775-793. doi:10.1016/j.ejor.2013.12.033.

Dijkstra, E.W., 1959. A note on two problems in connexion with graphs. Numerische Mathematik 1(1), 269-271. doi:10.1007/BF01386390.

Duan, M., Qi, G., Wei, G., Guo, R., 2020a. Comprehending and Analyzing Multiday Trip-Chaining Patterns of Freight Vehicles Using a Multiscale Method with Prolonged Trajectory Data. Journal of Transportation Engineering Part a-Systems 146(8). doi:10.1061/jtepbs.0000392.

Duan, M.Y., Qi, G.Q., Wei, G., Guo, R.G., 2020b. Comprehending and Analyzing Multiday Trip-Chaining Patterns of Freight Vehicles Using a Multiscale Method with Prolonged Trajectory Data. Journal of Transportation Engineering Part a-Systems 146(8). doi:10.1061/jtepbs.0000392.

Ester, M., Kriegel, H.-P., Sander, J., Xu, X., 1996. A density-based algorithm for discovering clusters in large spatial databases with noise, Proceedings of the Second International Conference on Knowledge Discovery and Data Mining. AAAI Press, Portland, Oregon, pp. 226–231. 

Evgenikos, P., Yannis, G., Folla, K., Bauer, R., Machata, K., Brandstaetter, C., 2016. Characteristics and causes of heavy goods vehicles and buses accidents in Europe, In: Rafalski, L., Zofka, A. (Eds.), Transport Research Arena Tra2016, pp. 2158-2167. doi:10.1016/j.trpro.2016.05.231.

Feng, T., Arentze, T., Timmermans, H., 2012. Spatial Environmental Analysis on the Effects of a New Navigation System for Freight Transport. Procedia - Social and Behavioral Sciences 54, 589-597. doi:10.1016/j.sbspro.2012.09.776.

Gingerich, K., Maoh, H., Anderson, W., 2016. Classifying the purpose of stopped truck events: An application of entropy to GPS data. Transportation Research Part C-Emerging Technologies 64, 17-27. doi:10.1016/j.trc.2016.01.002.

Gonzalez-Calderon, C.A., Holguin-Veras, J., Amaya, J., Sanchez-Diaz, I., Sarmiento, I., 2021. Generalized noortman and van es' empty trips model. Transportation Research Part a-Policy and Practice 145, 260-268. doi:10.1016/j.tra.2021.01.005.

Gonzalez-Feliu, J., Sanchez-Diaz, I., 2019. The influence of aggregation level and category construction on estimation quality for freight trip generation models. Transportation Research Part E-Logistics and Transportation Review 121, 134-148. doi:10.1016/j.tre.2018.07.007.

Greaves, S.P., Figliozzi, M.A., 2008. Collecting Commercial Vehicle Tour Data with Passive Global Positioning System Technology Issues and Potential Applications. Transportation Research Record(2049), 158-166. doi:10.3141/2049-19.

Hadavi, S., Verlinde, S., Verbeke, W., Macharis, C., Guns, T., 2019. Monitoring Urban-Freight Transport Based on GPS Trajectories of Heavy-Goods Vehicles. Ieee Transactions on Intelligent Transportation Systems 20(10), 3747-3758. doi:10.1109/tits.2018.2880949.

Haklay, M., Weber, P., 2008. OpenStreetMap: User-Generated Street Maps. Ieee Pervasive Computing 7(4), 12-18. doi:10.1109/mprv.2008.80.

Hassan, L.A., Mahmassani, H.S., Chen, Y., 2020. Reinforcement learning framework for freight demand forecasting to support operational planning decisions. Transportation Research Part E-Logistics and Transportation Review 137. doi:10.1016/j.tre.2020.101926.

Hess, S., Quddus, M., Rieser-Schuessler, N., Daly, A., 2015. Developing advanced route choice models for heavy goods vehicles using GPS data. Transportation Research Part E-Logistics and Transportation Review 77, 29-44. doi:10.1016/j.tre.2015.01.010.

Hu, W., Dong, J., Hwang, B.-g., Ren, R., Chen, Z., 2019. A Scientometrics Review on City Logistics Literature: Research Trends, Advanced Theory and Practice. Sustainability 11(10). doi:10.3390/su11102724.

Huang, J., Wang, L., Tian, C., Zhang, F., Xu, C., 2014. Mining freight truck's trip patterns from GPS data. 17th International IEEE Conference on Intelligent Transportation Systems (ITSC). 1988-1994. doi:10.1109/ITSC.2014.6957996.

Hughes, S., Moreno, S., Yushimito, W.F., Huerta-Cánepa, G., 2019. Evaluation of machine learning methodologies to predict stop delivery times from GPS data. Transportation Research Part C: Emerging Technologies 109, 289-304. doi:10.1016/j.trc.2019.10.018.

Jiang, D.D., Wang, W.J., Shi, L., Song, H.B., 2020. A Compressive Sensing-Based Approach to End-to-End Network Traffic Reconstruction. Ieee Transactions on Network Science and Engineering 7(1), 507-519. doi:10.1109/tnse.2018.2877597.

Joubert, J.W., Axhausen, K.W., 2011. Inferring commercial vehicle activities in Gauteng, South Africa. Journal of Transport Geography 19(1), 115-124. doi:10.1016/j.jtrangeo.2009.11.005.

Kamali, M., Ermagun, A., Viswanathan, K., Pinjari, A.R., 2016. Deriving Truck Route Choice from Large GPS Data Streams. Transportation Research Record(2563), 62-70. doi:10.3141/2563-10.

Knight, I., Newton, W., 2008. Longer and/or longer and heavier goods vehicles (LHVs): a study of the likely effects if permitted in the UK. Citeseer.

Kocatepe, A., Ozkul, S., Ozguven, E.E., Sobanjo, J.O., Moses, R., 2020. The Value of Freight Accessibility: a Spatial Analysis in the Tampa Bay Region. Applied Spatial Analysis and Policy 13(2), 527-546. doi:10.1007/s12061-019-09314-6.

Kovac, I., Vuletic, A., Mlinaric, D., 2020. Environmental responsibility of Croatian road freight transport enterprises. International Journal of Retail \& Distribution Management 48(9), 1023-1035. doi:10.1108/ijrdm-07-2019-0248.

Laranjeiro, P.F., Merchan, D., Godoy, L.A., Giannotti, M., Yoshizaki, H.T.Y., Winkenbach, M., Cunha, C.B., 2019. Using GPS data to explore speed patterns and temporal fluctuations in urban logistics: The case of Sao Paulo, Brazil. Journal of Transport Geography 76, 114-129. doi:10.1016/j.jtrangeo.2019.03.003.

Le Pira, M., Marcucci, E., Gatta, V., Inturri, G., Ignaccolo, M., Pluchino, A., 2017. Integrating discrete choice models and agent-based models for ex-ante evaluation of stakeholder policy acceptability in urban freight transport. Research in Transportation Economics 64, 13-25. doi:10.1016/j.retrec.2017.08.002.

Leblanc, L.J., Morlok, E.K., Pierskalla, W.P., 1975. EFFICIENT APPROACH TO SOLVING ROAD NETWORK EQUILIBRIUM TRAFFIC ASSIGNMENT PROBLEM. Transportation Research 9(5), 309-318. doi:10.1016/0041-1647(75)90030-1.

liu, F., Gao, Z., Janssens, D., Jia, B., Wets, G., Yang, Y., 2021. Identifying business activity-travel patterns based on GPS data. Transportation Research Part C: Emerging Technologies 128, 103136. doi:10.1016/j.trc.2021.103136.

Lorenz, M.O., 1905. Methods of Measuring the Concentration of Wealth. Publications of the American Statistical Association 9(70), 209-219. doi:10.1080/15225437.1905.10503443.

Louail, T., Lenormand, M., Cantu Ros, O.G., Picornell, M., Herranz, R., Frias-Martinez, E., Ramasco, J.J., Barthelemy, M., 2014. From mobile phone data to the spatial structure of cities. Scientific Reports 4. doi:10.1038/srep05276.

Luong, T.D., Tahlyan, D., Pinjari, A.R., 2018. Comprehensive Exploratory Analysis of Truck Route Choice Diversity in Florida. Transportation Research Record 2672(9), 152-163. doi:10.1177/0361198118784175.

Ma, X., McCormack, E.D., Wang, Y., 2011. Processing Commercial Global Positioning System Data to Develop a Web-Based Truck Performance Measures Program. Transportation Research Record(2246), 92-100. doi:10.3141/2246-12.

McCabe, S., Kwan, H., Roorda, M.J., 2013. COMPARING GPS AND NON-GPS SURVEY METHODS FOR COLLECTING URBAN GOODS AND SERVICE MOVEMENTS. International Journal of Transport Economics 40(2), 183-205. 

McCormack, E., Hallenbeck, M.E., Trb, 2006. ITS devices used to collect truck data for performance benchmarks, National, State, and Freight Data Issues and Asset Management, pp. 43-50. 

McCormack, E., Ma, X., Klocow, C., Curreri, A., Wright, D., 2010. Developing a GPS-based truck freight performance measure platform. 

McCormack, E., Zhao, W.J., Dailey, D.J., Ieee, 2012. GPS Tracking Of Freight Vehicles To Identify And Classify Bottlenecks, 2012 15th International Ieee Conference on Intelligent Transportation Systems, pp. 1245-1249. 

Moshref-Javadi, M., Lee, S., Winkenbach, M., 2020. Design and evaluation of a multi-trip delivery model with truck and drones. Transportation Research Part E-Logistics and Transportation Review 136. doi:10.1016/j.tre.2020.101887.

Oka, H., Hagino, Y., Kenmochi, T., Tani, R., Nishi, R., Endo, K., Fukuda, D., 2019. Predicting travel pattern changes of freight trucks in the Tokyo Metropolitan area based on the latest large-scale urban freight survey and route choice modeling. Transportation Research Part E-Logistics and Transportation Review 129, 305-324. doi:10.1016/j.tre.2017.12.011.

Pani, A., Sahu, P.K., 2019. Modelling non-response in establishment-based freight surveys: A sampling tool for statewide freight data collection in middle-income countries. Transport Policy. doi:10.1016/j.tranpol.2019.10.011.

Papadopoulos, A.-A., Kordonis, I., Dessouky, M.M., Ioannou, P.A., 2021. Personalized Pareto-improving pricing-and-routing schemes for near-optimum freight routing: An alternative approach to congestion pricing. Transportation Research Part C-Emerging Technologies 125. doi:10.1016/j.trc.2021.103004.

Parzen, E., 1962. On Estimation of a Probability Density Function and Mode. Ann. Math. Statist. 33(3), 1065-1076. doi:10.1214/aoms/1177704472.

Perez-Martinez, P.J., Andrade, M.d.F., de Miranda, R.M., 2017. Heavy truck restrictions and air quality implications in Sao Paulo, Brazil. Journal of Environmental Management 202, 55-68. doi:10.1016/j.jenvman.2017.07.022.

Perez-Martinez, P.J., Vassallo-Magro, J.M., 2013. Changes in the external costs of freight surface transport In Spain. Research in Transportation Economics 42, 61-76. doi:10.1016/j.retrec.2012.11.006.

Phan, M.H., Kim, K.H., 2016. Collaborative truck scheduling and appointments for trucking companies and container terminals. Transportation Research Part B-Methodological 86, 37-50. doi:10.1016/j.trb.2016.01.006.

Pluvinet, P., Gonzalez-Feliu, J., Ambrosini, C., 2012. GPS data analysis for understanding urban goods movement, In: Taniguchi, E., Thompson, R.G. (Eds.), Seventh International Conference on City Logistics, pp. 450-462. doi:10.1016/j.sbspro.2012.03.121.

Rosenblatt, M., 1956. Remarks on Some Nonparametric Estimates of a Density Function. Ann. Math. Statist. 27(3), 832-837. doi:10.1214/aoms/1177728190.

Sakai, T., Kawamura, K., Hyodo, T., 2017. Logistics Chain Modeling for Urban Freight Pairing Truck Trip Ends with Logistics Facilities. Transportation Research Record(2609), 55-66. doi:10.3141/2609-07.

Sakai, T., Kawamura, K., Hyodo, T., 2019. Evaluation of the spatial pattern of logistics facilities using urban logistics land-use and traffic simulator. Journal of Transport Geography 74, 145-160. doi:10.1016/j.jtrangeo.2018.10.011.

Sharma, S., Shelton, J., Valdez, G., Warner, J., 2020. Identifying optimal Truck freight management strategies through urban areas: Case study of major freight corridor near US-Mexico border. Research in Transportation Business \& Management, 100582. doi:10.1016/j.rtbm.2020.100582.

Shi, K.F., Chang, Z.J., Chen, Z.Q., Wu, J.P., Yu, B.L., 2020. Identifying and evaluating poverty using multisource remote sensing and point of interest (POI) data: A case study of Chongqing, China. J. Clean Prod. 255. doi:10.1016/j.jclepro.2020.120245.

Siripirote, T., Sumalee, A., Ho, H.W., 2020. Statistical estimation of freight activity analytics from Global Positioning System data of trucks. Transportation Research Part E-Logistics and Transportation Review 140. doi:10.1016/j.tre.2020.101986.

Sørensen, T., 1948. A Method of Establishing Groups of Equal Amplitude in Plant Sociology Based on Similarity of Species Content and Ist Application to Analyses of the Vegetation on Danish Commons. I kommission hos E. Munksgaard.

Stevens, C.J., Greif, A., Bouma, D., 2020. Do companies care about sustainable land governance? An empirical assessment of company land policies. International Journal of Sustainable Development and World Ecology 27(4), 334-348. doi:10.1080/13504509.2019.1701582.

Sun, L., Chen, J., Li, Q., Huang, D., 2020. Dramatic uneven urbanization of large cities throughout the world in recent decades. Nature Communications 11(1). doi:10.1038/s41467-020-19158-1.

Thakur, A., Pinjari, A.R., Zanjani, A.B., Short, J., Mysore, V., Tabatabaee, S.F., 2015. Development of Algorithms to Convert Large Streams of Truck GPS Data into Truck Trips. Transportation Research Record(2529), 66-73. doi:10.3141/2529-07.

Toilier, F., Serouge, M., Routhier, J.-L., Patier, D., Gardrat, M., 2016. How can Urban Goods Movements be Surveyed in a Megacity? The Case of the Paris Region. Transportation Research Procedia 12, 570-583. doi:10.1016/j.trpro.2016.02.012.

Velickovic, M., Stojanovic, D., Nikolicic, S., Maslaric, M., 2018. DIFFERENT URBAN CONSOLIDATION CENTRE SCENARIOS: IMPACT ON EXTERNAL COSTS OF LAST-MILE DELIVERIES. Transport 33(4), 948-958. doi:10.3846/16484142.2017.1350995.

Wang, H., Meng, Q., Zhang, X., 2020a. Multiple equilibrium behaviors of auto travellers and a freight carrier under the cordon-based large-truck restriction regulation. Transportation Research Part E-Logistics and Transportation Review 134. doi:10.1016/j.tre.2019.101829.

Wang, J., Morton, Y.T., 2015. High-Latitude Ionospheric Irregularity Drift Velocity Estimation Using Spaced GPS Receiver Carrier Phase Time-Frequency Analysis. Ieee Transactions on Geoscience and Remote Sensing 53(11), 6099-6113. doi:10.1109/tgrs.2015.2432014.

Wang, L., Wang, K., Zhang, J.J., Zhang, D., Wu, X., Zhang, L.J., 2020b. Multiple objective-oriented land supply for sustainable transportation: A perspective from industrial dependence, dominance and restrictions of 127 cities in the Yangtze River Economic Belt of China. Land Use Policy 99. doi:10.1016/j.landusepol.2020.105069.

Wang, X.K., Zhang, D.P., 2017. Truck freight demand elasticity with respect to tolls in New York State. Transportation Research Part a-Policy and Practice 101, 51-60. doi:10.1016/j.tra.2017.04.035.

Wang, Y., Yang, D., 2018. Impacts of Freight Transport on PM2.5 Concentrations in China: A Spatial Dynamic Panel Analysis. Sustainability 10(8). doi:10.3390/su10082865.

Wang, Z.Y., Bai, Y.F., Zhu, R., Wang, Y.L., Wu, B., Wang, Y.H., 2019. Impact Analysis of Extra Traffic Induced by Project Construction during Planned Special Events. Transportation Research Record 2673(7), 402-412. doi:10.1177/0361198119840346.

Yan, X.-Y., Wang, W.-X., Gao, Z.-Y., Lai, Y.-C., 2017. Universal model of individual and population mobility on diverse spatial scales. Nature Communications 8. doi:10.1038/s41467-017-01892-8.

Yang, X., Sun, Z., Ban, X.J., Holguin-Veras, J., 2014. Urban Freight Delivery Stop Identification with GPS Data. Transportation Research Record(2411), 55-61. doi:10.3141/2411-07.

Yang, X.C., Ye, T.T., Zhao, N.Z., Chen, Q., Yue, W.Z., Qi, J.G., Zeng, B., Jia, P., 2019. Population Mapping with Multisensor Remote Sensing Images and Point-Of-Interest Data. Remote Sensing 11(5). doi:10.3390/rs11050574.

Zanjani, A.B., Pinjari, A.R., Kamali, M., Thakur, A., Short, J., Mysore, V., Tabatabaee, S.F., 2015. Estimation of Statewide Origin-Destination Truck Flows from Large Streams of GPS Data Application for Florida Statewide Model. Transportation Research Record(2494), 87-96. doi:10.3141/2494-10.

Zhang, L., Hao, J., Ji, X., Liu, L., 2019. Research on the Complex Characteristics of Freight Transportation from a Multiscale Perspective Using Freight Vehicle Trip Data. Sustainability 11(7). doi:10.3390/su11071897.

Zhao, L., Song, Y.J., Zhang, C., Liu, Y., Wang, P., Lin, T., Deng, M., Li, H.F., 2020. T-GCN: A Temporal Graph Convolutional Network for Traffic Prediction. Ieee Transactions on Intelligent Transportation Systems 21(9), 3848-3858. doi:10.1109/tits.2019.2935152.

Zhen, L., Xia, J., Huang, L., Wu, Y.W., 2020. Bus tour-based routing and truck deployment for small-package shipping companies. Transportation Research Part E-Logistics and Transportation Review 136. doi:10.1016/j.tre.2020.101889.

\end{document}